\documentclass[a4paper,11pt]{article}
\usepackage{amsfonts,amsthm,amsmath,amssymb,graphicx,hyperref}

\def\Tr{{\rm Tr }}

\newcommand{\be}{\begin{equation}}
\newcommand{\bea}{\begin{eqnarray}}
\newcommand{\ee}{\end{equation}}
\newcommand{\eea}{\end{eqnarray}}

\newcommand{\bes}{\begin{equation*}}
\newcommand{\ees}{\end{equation*}}
\newcommand{\beas}{\begin{eqnarray*}}
\newcommand{\eeas}{\end{eqnarray*}}

\newcommand{\bmat}{\begin{bmatrix}}
\newcommand{\emat}{\end{bmatrix}}

\begin{document}

\begin{titlepage}

\thispagestyle{empty}

\begin{flushright}
YITP-19-45

\end{flushright}

\bigskip

\begin{center}
\noindent{{\Large \textbf {Chaos and Complexity in Quantum Mechanics}}}\\
\vspace{1cm}
	
Tibra Ali$^{(a)}$ \footnote{tali@perimeterinstitute.ca}, Arpan Bhattacharyya$^{(c),(b)}$   \footnote{abhattacharyya@iitgn.ac.in},  
S. Shajidul Haque$^{(d)}$\footnote{shajid.haque@uwindsor.ca},\\ Eugene H. Kim$^{(e)}$\footnote{ehkim@uwindsor.ca}, Nathan Moynihan$^{(d)}$\footnote{nathanmoynihan@gmail.com}, Jeff Murugan $^{(d)}$\footnote{jeff.murugan@uct.ac.za} 
 \\ ~~~~\\

$ {}^{(a)}$ {\it Perimeter Institute, \\
\it 31 Caroline Street North\\
 Waterloo, Ontario, Canada, N2L 2Y5 \\
}	
       
       ${}^{(c)}${\it Indian Institute of Technology,
Gandhinagar,\\ \it Gujarat 382355, India }

$ {}^{(b)}$ {\it Center for Gravitational Physics, \\
\it Yukawa Institute for Theoretical Physics (YITP), Kyoto University, \\
\it Kitashirakawa Oiwakecho, Sakyo-ku, Kyoto 606-8502, Japan\\
}	

$ {}^{(d)}$ {\it{The Laboratory for Quantum Gravity \& Strings,\\
	Department of Mathematics \& Applied Mathematics,\\
	University of Cape Town,\\
	Private Bag, Rondebosch, 7701, South Africa}\\ 
}

$ {}^{(e)}$ {\it Department of Physics, University of Windsor, \\
\it 401 Sunset Avenue\\
Windsor, Ontario, Canada, N9B 3P4\\
}	

\vskip 2em
\end{center}

\begin{abstract}
\vspace{-0.75cm}

\end{abstract}
We propose a new diagnostic for quantum chaos. We show that time evolution of complexity for a particular type of target state can provide equivalent information about the classical Lyapunov exponent and scrambling time as out-of-time-order correlators. Moreover, for systems that can be switched from a regular to unstable (chaotic) regime by a tuning of the coupling constant of the interaction Hamiltonian, we find that the complexity defines a new time scale. We interpret this time scale as recording when the system makes the transition from regular to chaotic behaviour. 

\end{titlepage}
\newpage
%\tableofcontents

%%%%%%%%%%%%%%%%%%%%%%%%%%%%%%%%%%%%%%%%%%%%%%%%%%%%%%%%%%%%%%%%%%%%%%%%%%%
%%%%%%%%%%%%%%%%%%%%%%%%%%%%%%%%%%%%%%%%%%%%%%%%%%%%%%%%%%%%%%%%%%%%%%%%%%%
%%%%%%%%%%%%%%%%%%%%%%%%%%%%%%%%%%%%%%%%%%%%%%%%%%%%%%%%%%%%%%%%%%%%%%%%%%%
%%%%%%%%%%%%%%%%%%%%%%%%%%%%%%%%%%%%%%%%%%%%%%%%%%%%%%%%%%%%%%%%%%%%%%%%%%%

\newpage
\section{Introduction}
Quantum chaos is intrinsically difficult to characterize. Consequently, a precise definition of quantum chaos in many-body 
systems remains elusive and our understanding of the dynamics of quantum chaotic systems is still inadequate.  
This lack of understanding is at the heart of a number of open questions in theoretical physics such as thermalization and transport in quantum many-body systems, and black hole information loss. It has also precipitated the renewed interest in quantum chaos from various branches of physics from condensed matter physics to quantum gravity \cite{thesis}.\\

Chaotic classical systems on the other hand are characterised by their {\it sensitive dependence on initial conditions} --- two copies of such a system, prepared in nearly identical initial states (namely, two distinct points in phase space, separated by a very small distance), will evolve over time into widely separated configurations. More precisely, the distance between the two points in phase space grows as $\exp(\lambda_L\, t)$, where $\lambda_L$ is the system's largest Lyapunov exponent \cite{l}. This does not happen in quantum mechanics --- two nearly identical states, {\it i.e.} states with a large initial overlap, remain nearly identical for all time (as their overlap is constant under unitary evolution). It has been argued \cite{peres,pastawski} that a quantum analog of ``sensitive dependence on initial conditions" is to consider evolving identical states with slightly different Hamiltonians, $\hat{H}$ and $\hat{H} + \delta \hat{H}$. If $\hat{H}$ is the quantization of a (classically) chaotic Hamiltonian, the states will evolve into two different states whose inner product decays exponentially in time.\\

Traditionally, chaos in quantum systems has been identified by comparison with results from random matrix theory (RMT)\cite{rc}. Recently however, other diagnostics have been proposed to probe chaotic quantum systems \cite{nd, nd1,nd2}. One such diagnostic is out-of-time-order correlators (OTOCs) \cite{ot, ot1} from which both the (classical) Lyapunov exponent  as well as the scrambling time \cite{sc,sc1,sc2} may be extracted.  However, recent work in mass-deformed SYK models \cite{Nosaka:2018iat} have revealed some tension between the OTOC and RMT diagnostics that arise, in part, through the nature of the probes. The OTOC captures early-time quantum mechanical features of the system while RMT diagnostics typically capture late-time statistical features. Evidently, a deeper understanding of probes of quantum chaos is required. In this light, it is interesting therefore to ask whether one can characterize chaos in quantum systems using quantum information-theoretic measures\footnote{Some progress in this direction was made in \cite{Miyaji:2016fse, qchaos2,qchaos3,qchaos4, qchaos,qchaos1,rubbish}.}. In this work, we propose a new diagnostic/probe of quantum chaos using the notion of {\sl circuit complexity} \cite{jm, jma, jmb, jmba, jmc, jmca, me, me1, me2, meen,club, cluba,clubb}, adopting Nielsen's geometric approach \cite{Nielsen1,Nielsen2, Nielsen3}.  More specifically, we study the circuit complexity of a particular target state obtained from a reference state by performing a forward evolution followed by a backward evolution with slightly different Hamiltonians. Then we demonstrate how this enables one to probe/characterize chaotic quantum systems, giving information beyond what is contained in the OTOC. Note that instead of using the complexity for a target state that is forward and then backward evolved from a reference state as mentioned above, one can as well study the complexity of a different circuit where both the target and reference states are obtained from time evolution (once) by applying slightly different Hamiltonians from some common state. For circuit complexity from the correlation matrix method that we will be using in this paper, the authors of \cite{me1} concretely showed that the time evolution of complexity in these two scenarios is identical. \\ %In that sense, we are basically studying the sensitivity of the geodesics (particularly geodesic distances) in the space of the unitaries.\\

To establish out testing method, we consider a simple, exactly solvable system --- the inverted oscillator, described by the Hamiltonian 
$H = p^2/2 - \omega^2 x^2/2$ \cite{barton} --- which captures the exponential sensitivity 
to initial conditions exhibited by chaotic systems \cite{zurek}.
Classically, the inverted oscillator has an unstable fixed point at $(x=0, p=0)$; a particle accelerates 
exponentially away from the fixed point when perturbed. Though the phase-space volume of the inverted 
oscillator is unbounded, our results are relevant to systems with a bounded phase space in that such a system would be described by an inverted oscillator up to a certain time. The two systems produce the same results over the time of interest (but would not be analytically solvable beyond that time).
In what follows, we include the analysis for a regular oscillator as a reference for what arises in a non-chaotic system and explore also a many-body system (quantum field theory) where the inverted oscillator appears. It is worth noting the inverted quantum oscillator is not just a toy model; it has been realized experimentally \cite{invertedexp} and has even played a role in mathematics, in attacking the Riemann hypothesis \cite{riemann}. It also provides important insights into the bound of the Lyapunov exponents \cite{Morita}.\\

The rest of the paper is organized as follows.
In Section 2, we present the model and states considered in this work. In Section 3, we review the 
ideas behind circuit complexity, and compute the circuit complexity for our system.
Sections 4 demonstrate how quantum chaos can be detected and quantified using
circuit complexity while Section 5 discusses the OTOC and its relation to the results obtained from 
the circuit complexity.  In Section 6, we discuss a many-body system (quantum field theory) 
where the inverted oscillator arises.  Finally, we summarize and present concluding remarks in Section 7.

%%%%%%%%%%%%%%%%%%%%%%%%%%%%%%%%%%%%%%%%%%%%%%%%%%%%%%%%%%%%%%%%%%%%%%%%%%%
%%%%%%%%%%%%%%%%%%%%%%%%%%%%%%%%%%%%%%%%%%%%%%%%%%%%%%%%%%%%%%%%%%%%%%%%%%%
%%%%%%%%%%%%%%%%%%%%%%%%%%%%%%%%%%%%%%%%%%%%%%%%%%%%%%%%%%%%%%%%%%%%%%%%%%%
%%%%%%%%%%%%%%%%%%%%%%%%%%%%%%%%%%%%%%%%%%%%%%%%%%%%%%%%%%%%%%%%%%%%%%%%%%%

% \section{Our Model and Quench Protocol}
\section{The Model}

We are interested in comparing the complexity of a regular system with that of an unstable/chaotic system.
To that end, we consider the Hamiltonian
\begin{equation} \label{Ham}
 H = \frac{1}{2} p^2 + \frac{\Omega^2}{2} x^2
 \ \ {\rm where} \ \  \Omega^2 = m^2 - \lambda \ .
\end{equation}
For $\lambda < m^2$, equation (\ref{Ham}) describes a simple harmonic oscillator; for
$\lambda>m^2$, we have an inverted oscillator. The $\lambda = m^2$ case, of course describes a free particle.
Our inverted oscillator model can be understood as a short-time approximation for unstable/chaotic systems. In particular, this model captures the exponential sensitivity to initial conditions exhibited by chaotic systems. Let's start with the following state at $t=0$,
\be \label{1}
\psi(x,t=0)=\mathcal{N}(t=0)\exp \left (- \frac{\omega_{r} \, x^2}{2} \right),
\ee
where 
\be
\omega_r=m.
\ee 
Evolving this state in time by the Hamiltonian (\ref{Ham}) produces \cite{shankar}
\begin{align}
\begin{split} \label{2}
\psi (x,t) =\mathcal{N}(t) \, \exp \left (- \frac{ \omega(t) \, x^2}{2} \right) \ ,
\end{split}
\end{align}
where $\mathcal{N}(t)$ is the normalization factor and
\be  \label{Omega}
\omega(t)=\Omega \left (\frac{\Omega-i\,\omega_r \cot (\Omega\, t)}{\omega_r-i\,\Omega \cot (\Omega\, t)}\right).
\ee
We will be computing the complexity for this kind of time evolved state (\ref{2}) with respect to (\ref{1}) and 
\be
\omega(t=0)=\omega_r.
\ee

%%%%%%%%%%%%%%%%%%%%%%%%%%%%%%%%%%%%%%%%%%%%%%%%%%%%%%%%%%%%%%%%%%%%%%%%%%%
%%%%%%%%%%%%%%%%%%%%%%%%%%%%%%%%%%%%%%%%%%%%%%%%%%%%%%%%%%%%%%%%%%%%%%%%%%%
%%%%%%%%%%%%%%%%%%%%%%%%%%%%%%%%%%%%%%%%%%%%%%%%%%%%%%%%%%%%%%%%%%%%%%%%%%%
%%%%%%%%%%%%%%%%%%%%%%%%%%%%%%%%%%%%%%%%%%%%%%%%%%%%%%%%%%%%%%%%%%%%%%%%%%%

\section{Complexity from the Covariance Matrix}

We will start this section with a quick review of circuit complexity and then conclude with a computation of the circuit complexity for a single oscillator. For circuit complexity we will use the covariance matrix method. Note that a similar analysis can be done for circuit complexity from the full wave function. 
\subsection{Review of Circuit Complexity}
Here we will briefly sketch the outline of the computation of circuit complexity. Details of this can be found in \cite{jm, jmb}. We will highlight only the key formulae and interested readers are referred to \cite{jm,jmb} and citations thereof. The problem is simple enough to state; given a set of elementary gates and a reference state, we want to build the most efficient circuit that starts at the reference state and terminates at a specified target state. Formally,
\begin{align}
\begin{split} \label{statement}
|\psi_{\tau=1}\rangle=\tilde U(\tau=1)|\psi_{\tau=0}\rangle,
\end{split}
\end{align}
where \be \label{Unit}
\tilde U(\tau)= {\overleftarrow{\mathcal{P}}}\exp(i\int_{0}^{\tau }\, d\tau\, H(\tau)),
\ee 
is the unitary operator representing the quantum circuit, which takes the reference state $|\psi_{\tau=0}\rangle$ to the target state $|\psi_{\tau=1}\rangle$.
 $\tau$ parametrizes a path in the space of the unitaries and  given a particular basis (elementary gates) $M_I$, 
\begin{equation}
 H(\tau)= Y^{I}(\tau) M_{I} \ .
\nonumber
\end{equation}
In this context, the coefficients $\{ Y^{I}(\tau) \}$ are referred to as `control functions'. The path ordering in (\ref{Unit}) is necessary as all the $M_I$'s do not necessarily commute with each other. \\

Now, since the states under consideration (\ref{1}) and (\ref{2}) are Gaussian, they can be equivalently described by a \textit{Covariance matrix} as follows
\be
   G^{ab}= \langle \psi(x,t)|\xi^a\xi^b+\xi^b\xi^a|\psi(x,t) \rangle,
\ee
where $\xi=\{x ,p\}.$  This covariance matrix is typically a real symmetric matrix with unit determinant. We will always transform the reference covariance matrix such that \cite{jmca, me1}
\be
\tilde G^{\tau=0}= S\cdot G^{\tau=0}\cdot S^T
\ee
with $\tilde G^{\tau=0}$ an identity matrix and $S$ a real symmetric matrix whose transpose is denoted $S^T$. Similarly, the reference state will transform as
\be
\tilde G^{\tau=1}= S\cdot G^{\tau=1}\cdot S^{T}.
\ee
The unitary $\tilde U(\tau)$ acts on this transformed covariance matrix as,  
\be
\tilde G^{\tau=1}= \tilde U(\tau)\cdot \tilde G^{\tau=0}\cdot\tilde U^{-1}(\tau).
\ee
Next we define suitable \textit{cost function} $\mathcal{F}(\tilde U, \dot {\tilde U})$ and define \cite{jm,Nielsen1,Nielsen2,Nielsen3}
\be \label{cost}
\mathcal{C}(\tilde U)=\int_{0}^{1} \mathcal{F}(\tilde U,\dot{\tilde U})\, d\tau \ .
\ee
Minimizing this  cost functional gives us the optimal set of $Y^{I}(\tau)$, which in turn give us the most efficient circuit by minimizing the circuit depth. There are various possible choices for these  functions $\mathcal{F}(\tilde U,\dot{\tilde U}).$ For further details, we refer the reader to the extensive literature in \cite{jm, jma,jmb, Nielsen1, Nielsen2,Nielsen3}. In this paper, we will choose
\be
 \mathcal{F}_2(U,Y)=\sqrt{\sum_{I} (Y^{I})^2}.
 \ee
 For this choice, one can easily see that, after minimization the $\mathcal{C}(\tilde U)$ defined in (\ref{cost}) corresponds to the geodesic distance on the manifold of unitaries. Note also that we can reproduce our analysis done in the following sections with other choices of cost functional. We will, however, leave this for future work. 
 
 %%%%%%%%%%%%%%%%%%%%%%%%%%%%%%%%%%%%%%%%%%%%%%%%%%%%%%
\subsection{Circuit Complexity for a Single Oscillator}
 %%%%%%%%%%%%%%%%%%%%%%%%%%%%%%%%%%%%%%%%%%%%%%%%%%%%%%
 
For our case,  the covariance matrix corresponding  to target state (\ref{2}) will take the form,
\be
G^{\tau=1}=\left(
\begin{array}{cc}
 \frac{1}{ \text {Re}  (\omega(t))}& -\frac{\text {Im}  (\omega(t))}{\text {Re}  (\omega(t))} \\
-\frac{ \text {Im}  (\omega(t))}{\text {Re}  (\omega(t))} & \frac{|\omega(t) |^2}{\text {Re}  (\omega(t))} \\
\end{array}
\right),
\ee
where $\omega(t)$ is defined in (\ref{Omega}). For the reference state (\ref{1}) it will take the following form,
\be
G^{\tau=0}=\left(
\begin{array}{cc}
 \frac{1}{\omega_r}& 0 \\
0 & \omega_r \\
\end{array}
\right).
\ee
 Next we change the basis as follows
 \be
\tilde G^{\tau=1}= S\cdot G^{\tau=1}\cdot S^{T}, \qquad\tilde G^{\tau=0}= S\cdot G^{\tau=0}\cdot S^{T}, 
\ee
with \be S=\left(
\begin{array}{cc}
\sqrt{\omega^r} & 0\\
0& \frac{1}{\sqrt{\omega_r}}\\
\end{array}
\right),
\ee
such that $\tilde G^{\tau=0}= I $ is an identity matrix. For the case under study, the reference frequency $\omega_r$ is real. We will choose the following three generators,
\begin{align}
\begin{split}
M_{11}\rightarrow \frac{i}{2}( x \, p+p\, x),\quad M_{22}\rightarrow \frac{i}{2} x^2,\quad M_{33}\rightarrow \frac{i}{2} p^2.
\end{split}
\end{align}
These will serve as our elementary gates and satisfy the $SL(2,R)$ algebra. 
\begin{align}
\begin{split}
[M_{11}, M_{22}]=2\,M_{22},\quad[M_{11},M_{33}]=-2\,M_{33},\quad [M_{22},M_{33}]= M_{11}.
\end{split}
\end{align}
Next, if we parameterize the $\tilde U(\tau)$ as,
\begin{align}
\begin{split}\label{param}
\tilde U(\tau)=\left(
\begin{array}{cc}
\cos(\mu(\tau))\cosh(\rho(\tau))-\sin(\theta(\tau))\sinh(\rho(\tau))& -\sin(\mu(\tau))\cosh(\rho(\tau))+\cos(\theta(\tau))\sinh(\rho(\tau))\\
\sin(\mu(\tau))\cosh(\rho(\tau))+\cos(\theta(\tau))\sinh(\rho(\tau))&  \cos(\mu(\tau))\cosh(\rho(\tau))+\sin(\theta(\tau))\sinh(\rho(\tau))\\
\end{array}
\right).
\end{split}
\end{align} 
and set the boundary conditions as,
 \be
\tilde G^{\tau=1}=\tilde U(\tau=1)\cdot \tilde G^{\tau=0}\cdot\tilde U^{-1}(\tau=1),\qquad \tilde G^{\tau=0}=\tilde U(\tau=0)\cdot\tilde  G^{\tau=0}\cdot\tilde U^{-1}(\tau=0)\,,
\ee
we find that \cite{me1},
\be \label{boundary}
\{\cosh(2\rho(1)), \tan(\theta(1)+\mu(1)) \}=\left\{\frac{\omega_r^2+|\omega(t)|^2}{2\,\omega_r\,\text {Re}  (\omega(t))},\frac{\omega_r^2-|\omega(t)|^2}{2\,\omega_r\,\text {Im}  (\omega(t))}\right\},\quad
\{\rho(0), \theta(0)+\mu(0)\}=\{0, c\}.
\ee
Here $c$ is an arbitrary constant. For simplicity we choose $$\mu(\tau=1)=\mu(\tau=0)=0, \quad\theta(\tau=0)=\theta(\tau=1)=c=\tan^{-1}\left(\frac{\omega_r^2-|\omega(t)|^2}{2\,\omega_r\,\text {Im}  (\omega(t))}\right).$$ From (\ref{Unit}) we  have,
\be
Y^{I}=\Tr\left(\partial_{\tau}\tilde U(\tau)\cdot \tilde U(\tau)^{-1}\cdot (M^{I})^{T}\right) \ ,
\ee
where $\Tr \left(M^{I}. (M^{J})^{T}\right)=\delta^{IJ}$. Using this we can define the metric
\be
   ds^2= G_{IJ} dY^{I} dY^{* J},
\ee
where the $G_{IJ}=\frac{1}{2}\delta_{IJ}$ is known as a penalty factor. Given the form of $U(s)$ in (\ref{param}) we will have,
\be
ds^2=d\rho^2+\cosh(2\rho)\cosh^2\rho \,d\mu^2+\cosh(2\rho)\sinh^2\rho\, d\theta^2-\sinh(2\rho)^2\,d\mu\, d\theta,
\ee
and the complexity functional defined in (\ref{cost}) will take the form,
\be \label{cost1}
\mathcal{C}(\tilde U)=\int_{0}^{1}d\tau \sqrt{g_{ij}\dot x^{i}\dot x^{j}}.
\ee
The simplest solution for the geodesic is again a straight line on this geometry \cite{jm,me1}.
\be
\rho(\tau)= \rho(1)\, \tau.
\ee

 Evaluating (\ref{cost1}) we  simply get 
\begin{align}
\begin{split} \label{answ}
\mathcal{C}(\tilde U)= \rho(1)=\frac{1}{2}\left(\cosh^{-1}\left[\frac{\omega_r^2+|\omega(t)|^2}{2\,\omega_r\,\text {Re}  (\omega(t))}\right]\right) .
\end{split}
\end{align}
This is the geodesic distance in the space of $SL(2,R)$ unitaries with end points anchored at the two points determined the boundary conditions (\ref{boundary}).

\section{Quantifying Chaos using Complexity}

\noindent
The goal of this paper is to explore whether we can implement the notion of quantum circuit complexity as a diagnostic of a system's chaotic behaviour. Classically, chaos is diagnosed by studying trajectories in the phase space of some dynamical systems, a notion that is not well defined in quantum systems, essentially because of the uncertainty principle. It is important to keep in mind that when we speak of geodesics in the context of circuit complexity, we will mean trajectories defined on the space of unitaries.\\

Now we propose a new diagnostic for chaotic behaviour based on circuit complexity. We consider a target state $| \psi_2 \rangle$ obtained by evolving a reference state $| \psi_0 \rangle$ forward in time with $\hat{H}$ and then backward in time with $\hat{H}+ \delta \hat{H}$, 
\be
|\psi_2 \rangle = e^{i (\hat{H}+ \delta \hat{H}) t} e^{-i \hat{H} t} |\psi_0\rangle\,.
\ee
We would like to compute the complexity $\mathcal{\hat C}(\tilde U)$ of this target state $| \psi_2\rangle$ with 
respect to the reference state $| \psi_0\rangle$ \cite{me1}. For a chaotic quantum system, even if the two Hamiltonians $\hat{H}$ and $\hat{H} + \delta \hat{H}$
are arbitrarily close, $|\psi_2\rangle$ will be quite different from $|\psi_0\rangle.$\\

 \begin{figure}[ht] 
\centering
\scalebox{0.30}{\includegraphics{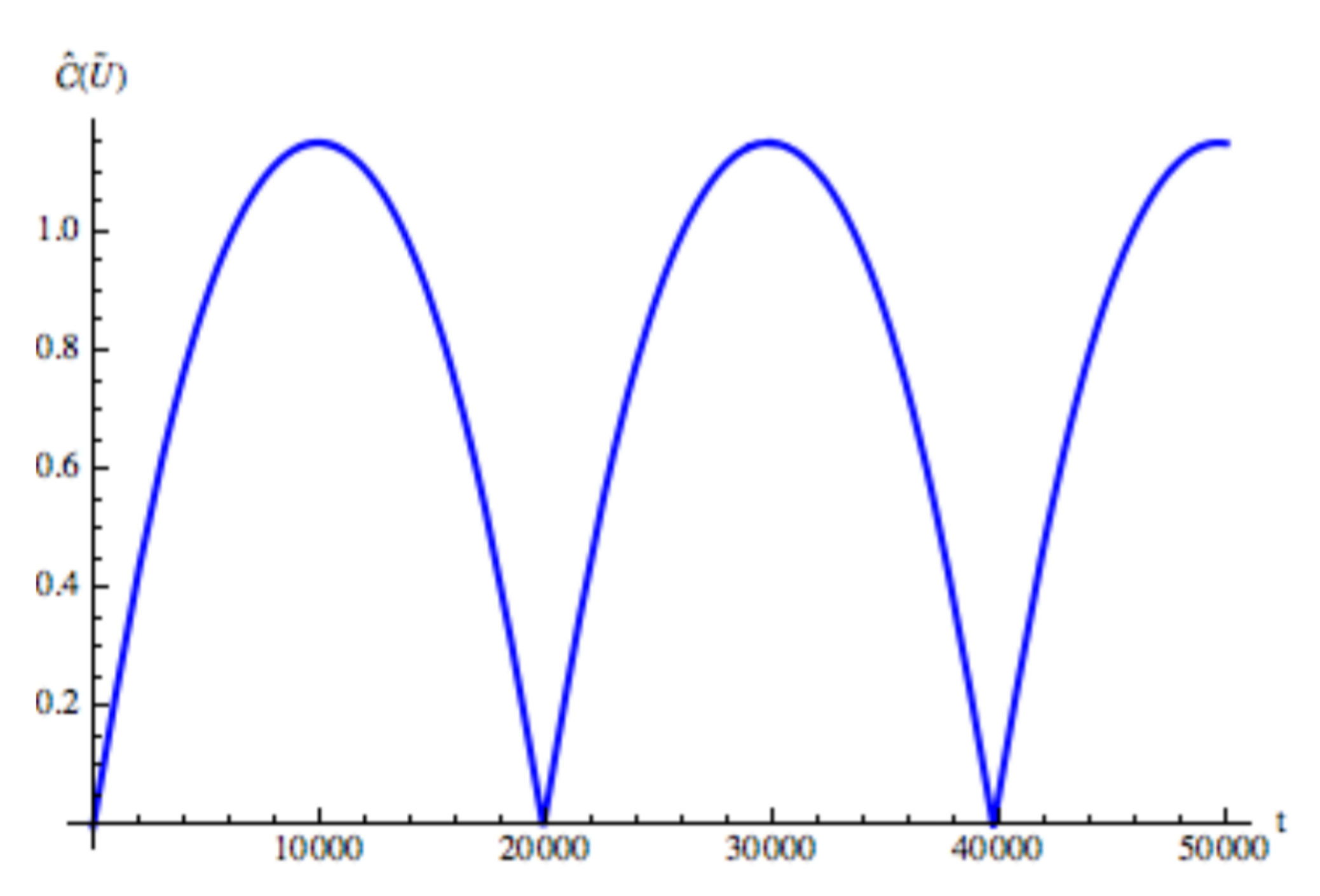}}
\caption{$\mathcal{\hat C}(\tilde U)$ vs time for Regular Oscillator ($m=1,\lambda=1.2,\delta\lambda=0.01$)}
\label{reg}
\end{figure}
\begin{figure}[ht] 
\centering
\scalebox{0.30}{\includegraphics{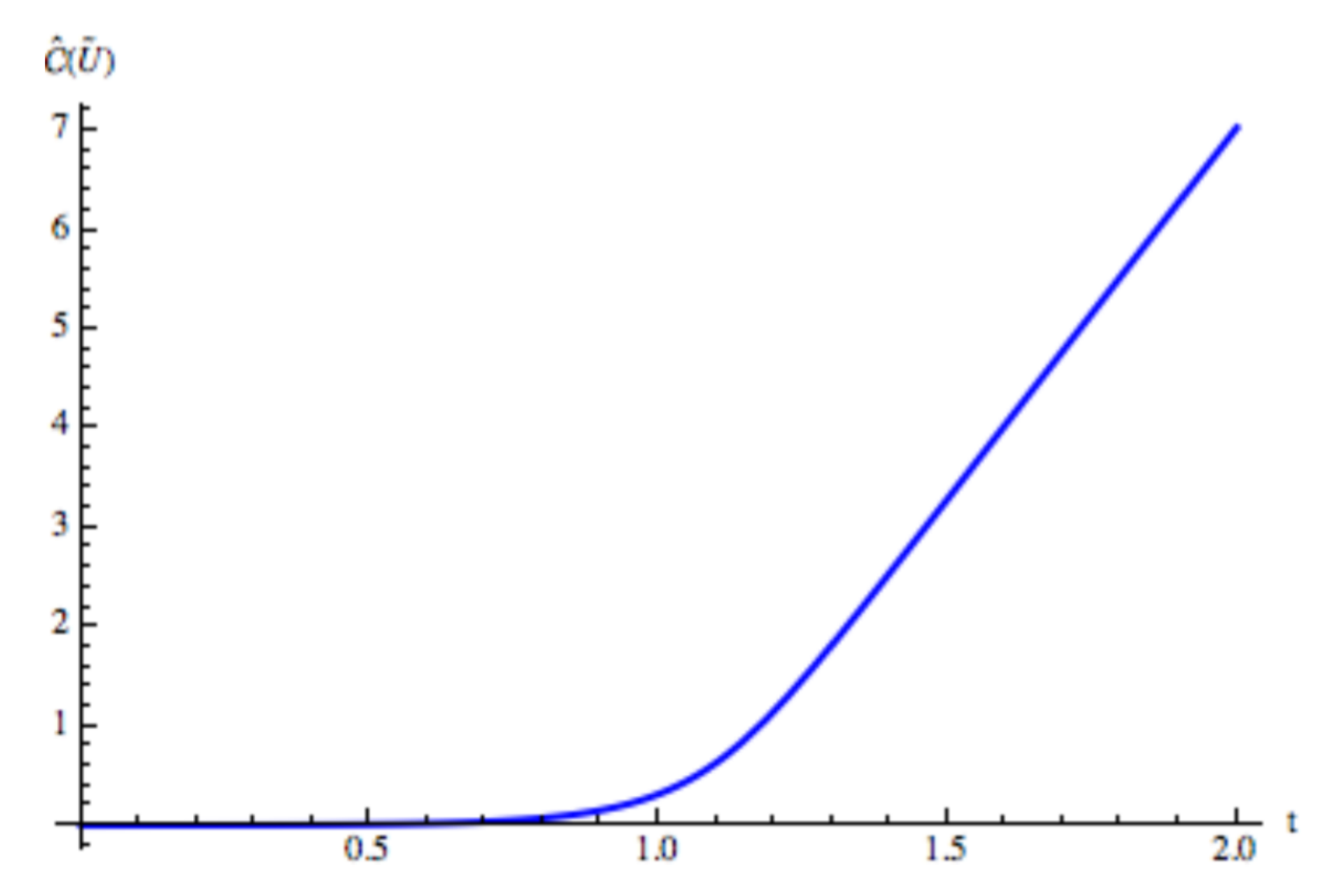} }
\caption{$\mathcal{\hat C}(\tilde U)$ vs time for Inverted Oscillator ($m =1, \lambda= 15, \delta\lambda =0.01$)}
\label{invert}
\end{figure}

\be
\mathcal{\hat C}(\tilde U) =\frac{1}{2}\Big(\cosh^{-1}\big[\frac{\omega_r^2+|\hat \omega(t)|^2}{2\,\omega_r\,\text {Re}  (\hat \omega(t))}\big]\Big),
\ee
where now
\be
\psi_2(x,t) =\mathcal{\hat N}(t) \exp \left[ -\frac{1}{2}\hat \omega(t) x ^2  \right] \,,
\ee
and
 \be \label{omedef} \hat \omega(t)=  i \ \Omega' \cot (\Omega' t)  + \frac{\Omega'^2}{  \sin^2 (\Omega' t) (\omega(t) + i \, \Omega' \cot (\Omega' t))}\,. 
\ee
%%%%%%%%
\begin{figure}[ht] 
\centering
\scalebox{0.30}{\includegraphics{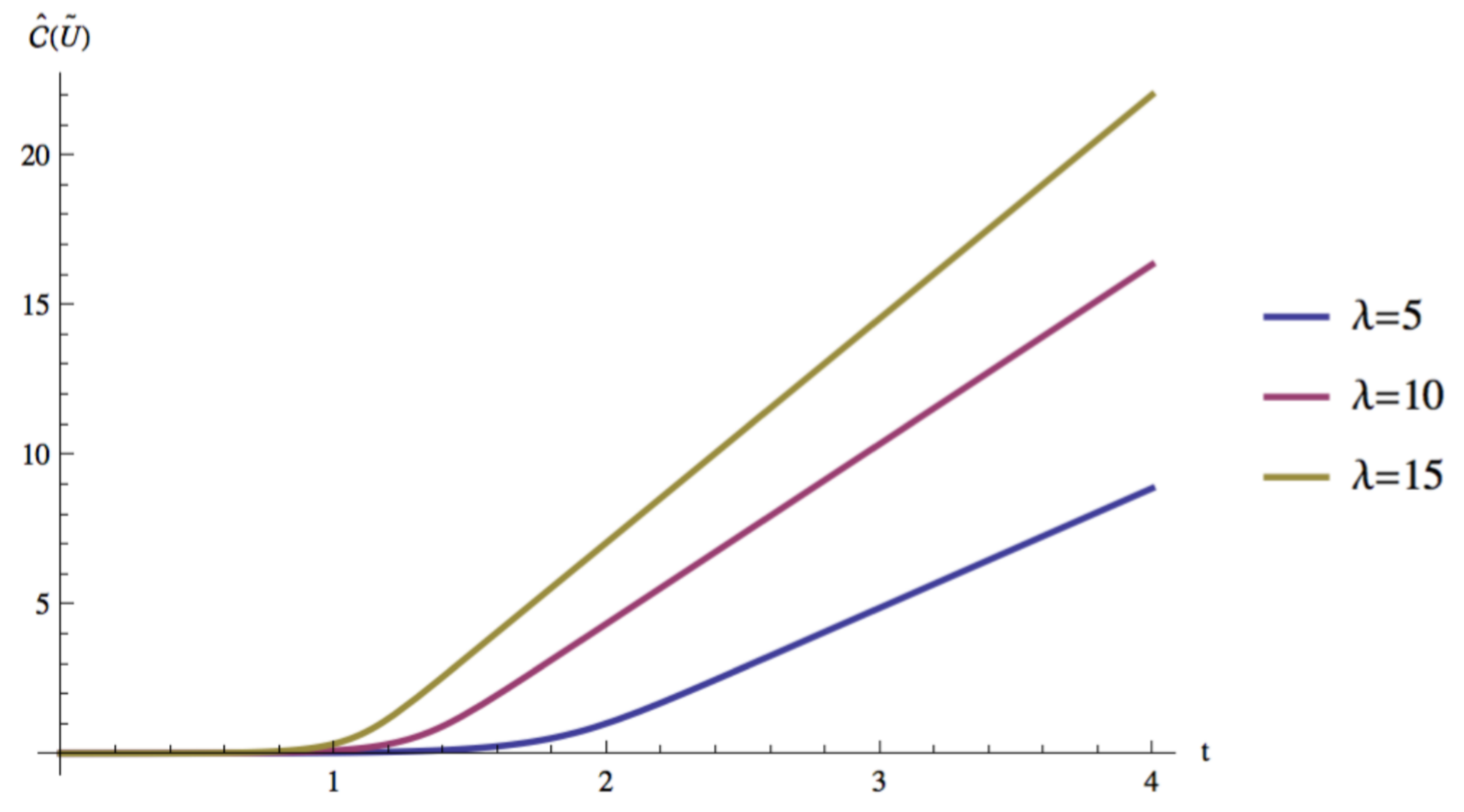} }
\caption{$\mathcal{\hat C}(\tilde U)$  vs time for different values of $\lambda$ (with $\delta\lambda=0.01,m=1$)}
\label{slope}
\end{figure}
%%%%%%%%%%
\begin{figure}[b] 
\centering
\scalebox{0.40}{\includegraphics{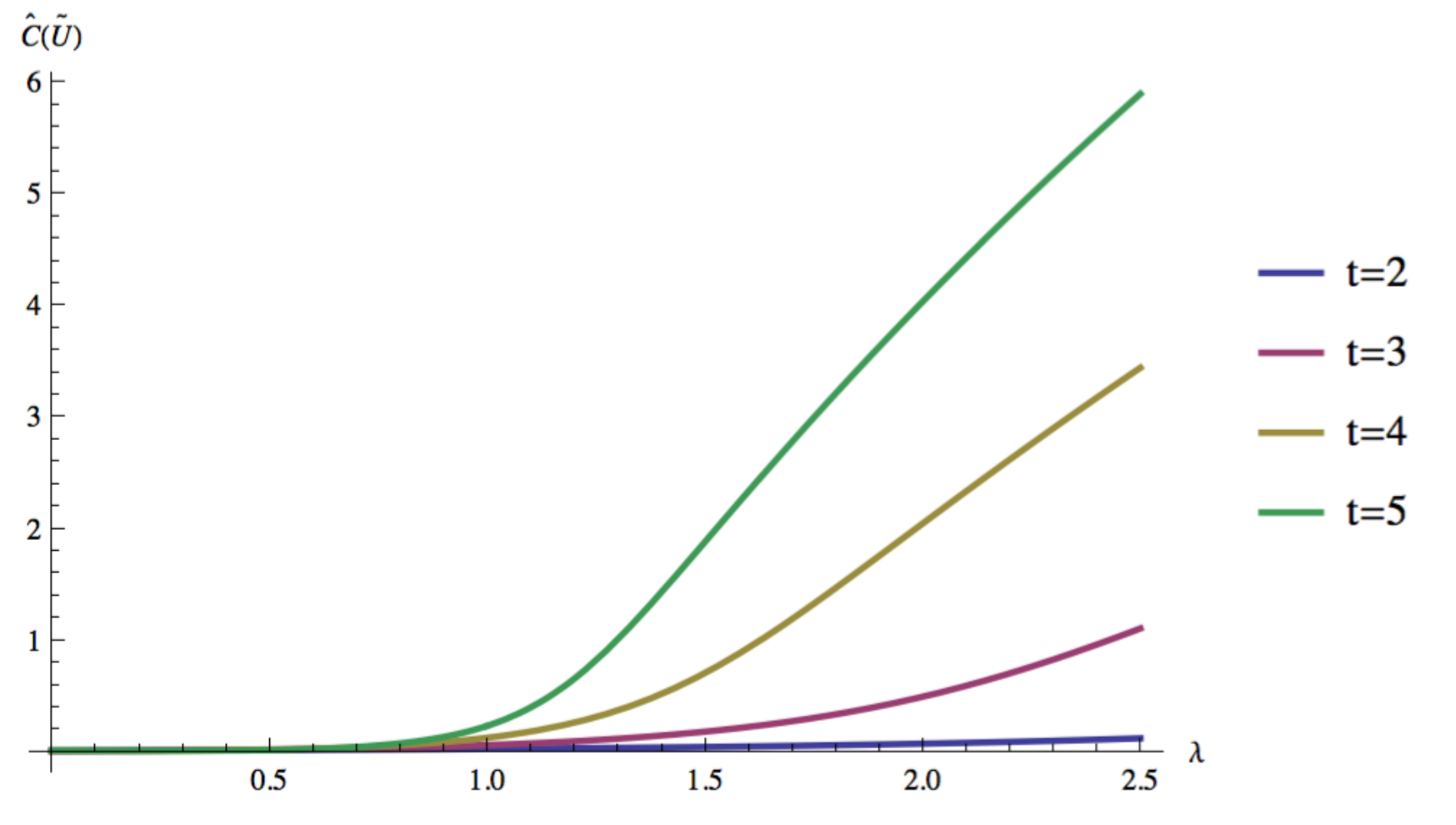} }
\scalebox{0.40}{\includegraphics{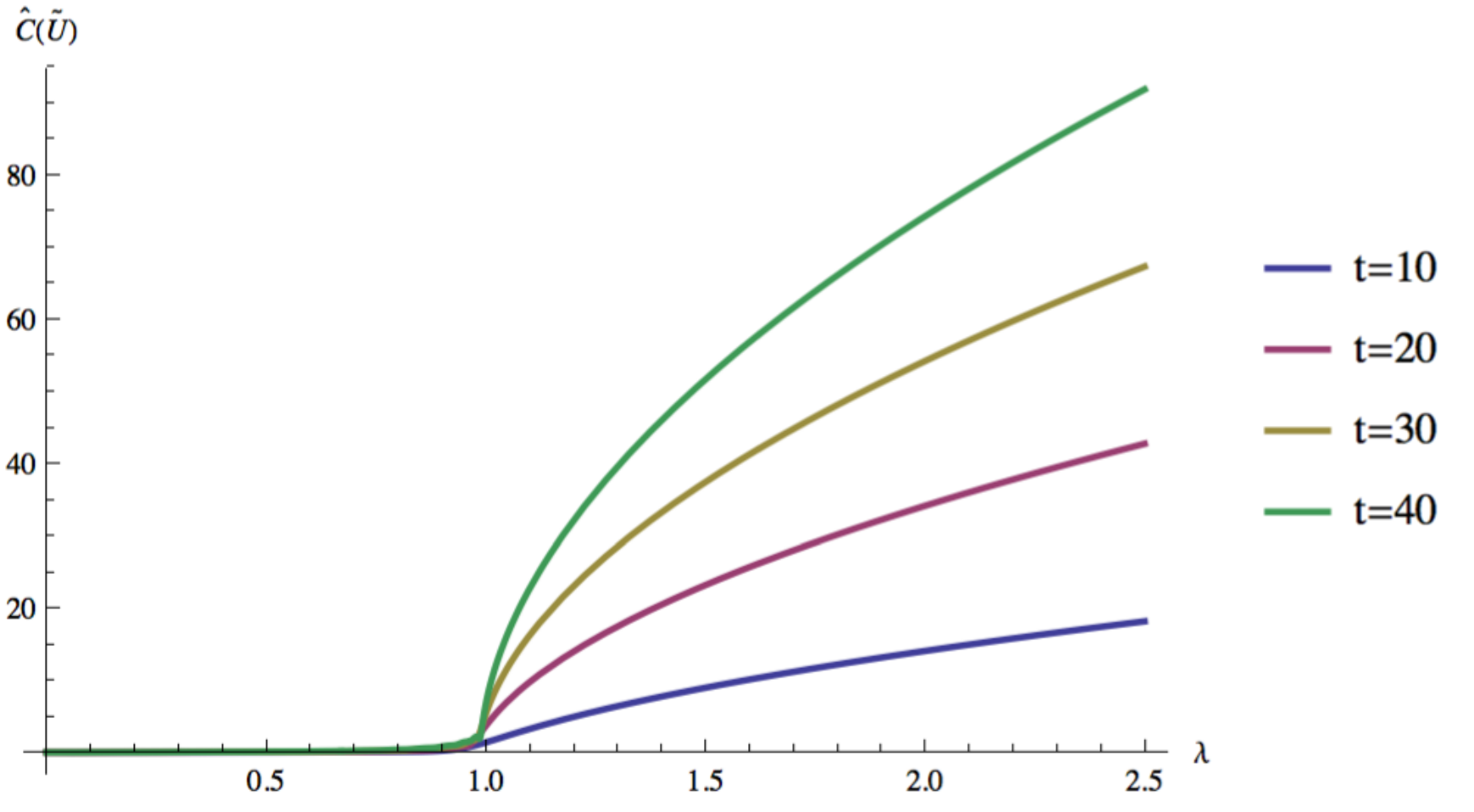}}
%\scalebox{0.40}{\includegraphics{LE1} }
\caption{Complexity vs $\lambda$ for different time. (a) In the left figure, for all $t$, $\hat{\cal C}(\tilde U) $ starts to increase before the critical of $\lambda$ namely $\lambda=m^2.$ (b) In the right figure, we can observe that near $t=40$ there is a sharp increase in $\hat{\cal C}(\tilde U)$ at the critical value of $\lambda$ ($\lambda=1$, for the choice of the parameter). We have set $\delta \lambda=0.01$ and $m=1$ for both the figures.}
\label{fig:time}
\end{figure}

\noindent
In this last expression, $\Omega'= \sqrt{m^2 - \lambda'}$ is the frequency associated with the Hamiltonian $H'=\frac{1}{2} p^2 + \frac{\Omega'^2}{2} x^2$ and $\lambda'=\lambda+\delta\lambda$ with $\delta\lambda$ very small. The time dependence of this complexity demonstrates that there is a clear qualitative difference between a regular oscillator and an inverted oscillator as evident from Fig.~(\ref{reg}) and Fig.~(\ref{invert}). For the regular oscillator we get oscillatory behaviour \cite{me1,me2, meen}; the complexity grows linearly for a very short period and reach to a saturation with some fluctuations. However, $\mathcal{\hat C}(\tilde U)$ for the inverted oscillator tells a completely different story. \\

The overall behaviour of $\mathcal{\hat C}(\tilde U)$ for the inverted oscillator appears to be some complicated monotonically growing function. However, a closer look at Fig.~(\ref{invert}), reveals that it takes a small amount of time for the complexity to pick up after which it displays a linear ramp with time. For a different choice of coupling ($\lambda >\lambda_c$) we get similar behaviour with different pick up time and slope  ($\phi$) for the linear ramp. These features are displayed for different values of the coupling in Fig.~(\ref{slope}). \\

\begin{figure}[ht] 
\centering
\scalebox{0.35}{\includegraphics{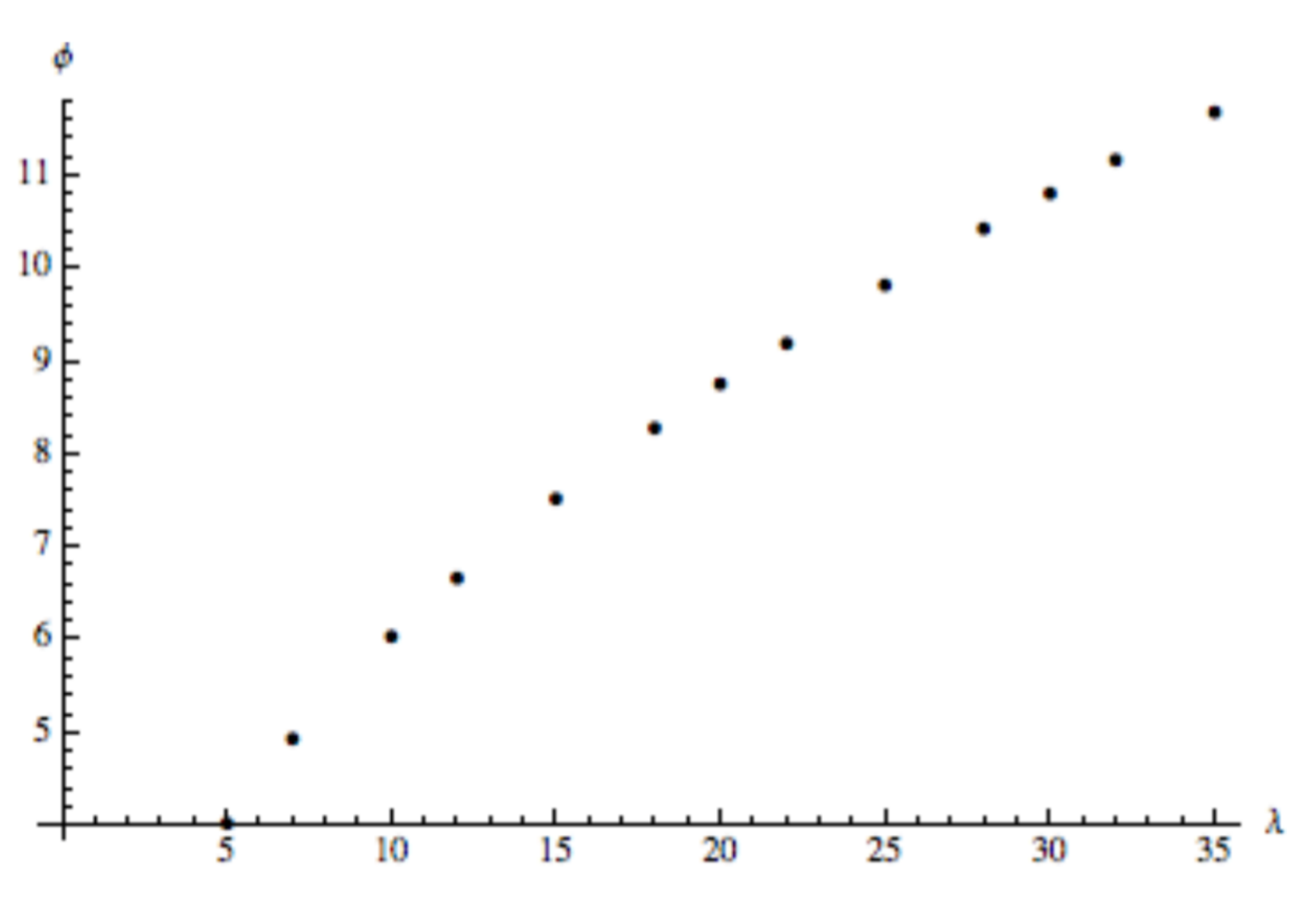} }
\scalebox{0.35}{\includegraphics{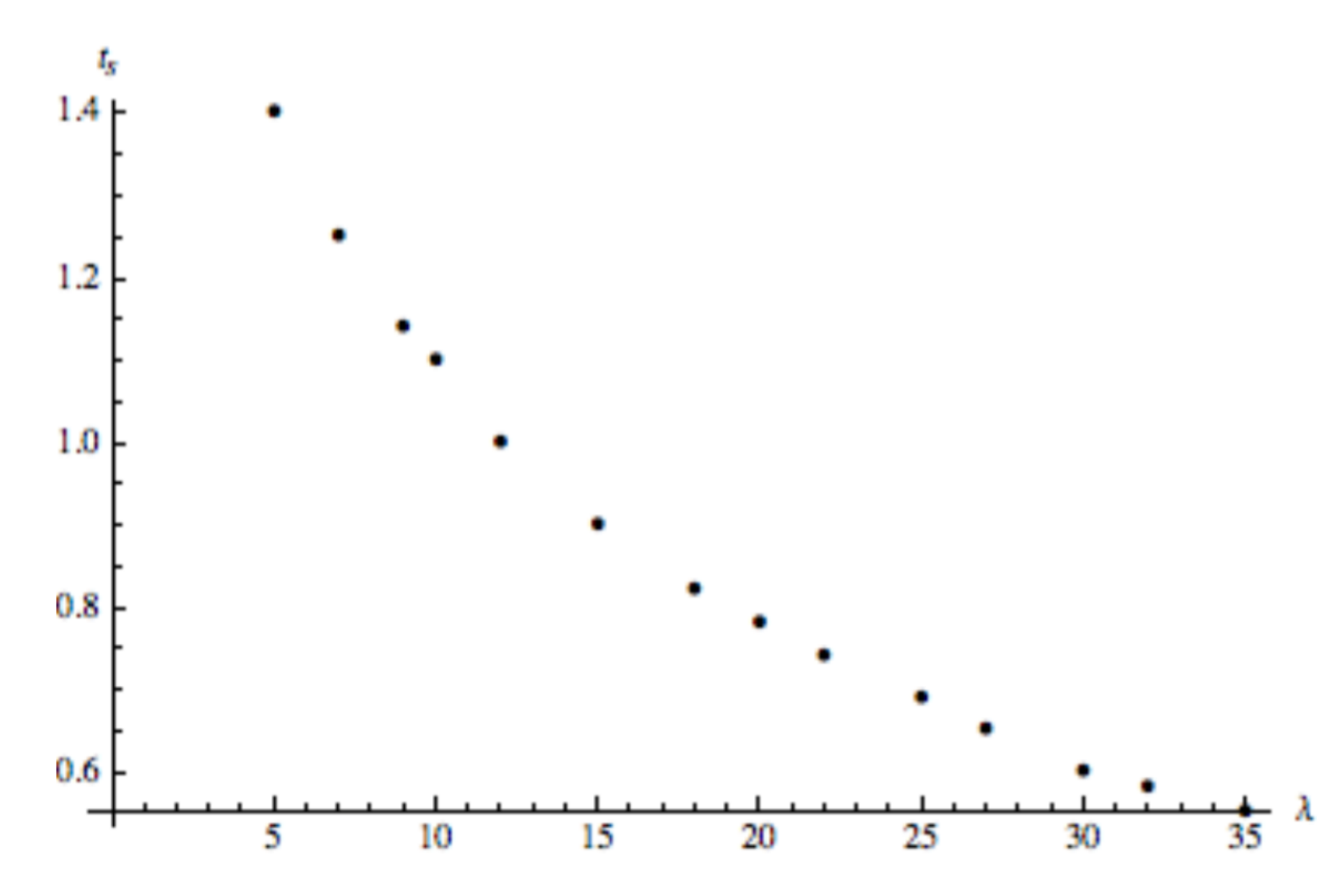} }
\caption{(a) Slope $\phi$ vs $\lambda$ ($\delta=0.01,m=1$), (b)$t_s$ vs $\lambda$($\delta=0.01,m=1$)}
\label{lyap3}
\end{figure}

As we increase $\lambda$ (beyond the critical value), we are in effect making the model more unstable and consequently from our very specific circuit model we expect a larger complexity and a smaller pick up time. Therefore, the slope and pick up time scale are natural candidates for measuring the unstable nature of the inverted oscillator. When we explore the slope $\phi$ of the linear region (as in Fig.~(\ref{slope})) for different values of  coupling $\lambda$ we find the behaviour shown in Fig.~(\ref{lyap3}). In the following section, we will argue that this slope is similar to the Lyapunov exponent. \\

%%%%%%
%%%%%%%%%%%%%%
%%%%%%%%%%%%%%

Note that the linear growth kicks in near a certain time $t=t_s$ (as in the Fig.~(\ref{invert})) which depends on the choice of the parameters, $\{m, \lambda\}$. We plot this pick up time as function of $\lambda$ in Fig.~(\ref{lyap3}b). We believe that this time scale is equivalent to the scrambling time which frequently appears in the chaos literature. One way to confirm this is to compute the out-of-time-order four point correlator (OTOC). The time when OTOC $ \sim e^{\Lambda (t-t_*)}$ becomes $  \mathcal{O}(1)$, is called the scrambling time. For this oscillator model (\ref{Ham}), we can show it analytically following \cite{Hashimoto:2017oit}. It is shown in the next section.\\

%%%%%%%%%%%%%
%%%%%%%%%%%%

Now we will study how this complexity changes with coupling $\lambda$ for a fixed time.  Fig.~(\ref{fig:time}) shows how the complexity changes with coupling $\lambda$ for various times. We denote $\lambda_c=m^2$ as the critical value of the $\lambda,$ after which the system becomes an inverted oscillator. We find that for smaller values of $t$ the complexity start to increase for $\lambda < \lambda_c.$ Only around $t=40$ the complexity sharply increases at $\lambda=\lambda_c.$ We call this time as the critical time $t_c.$ Fig.~(\ref{fig:time}) shows that, it takes a certain amount of time for the system to ``know" that it has become chaotic; $t_c$ marks when this occurs.\\

%%%%%%%%%%%%%
%%%%%%%%%%%%%

\begin{figure}[ht] 
\centering
\scalebox{0.30}{\includegraphics{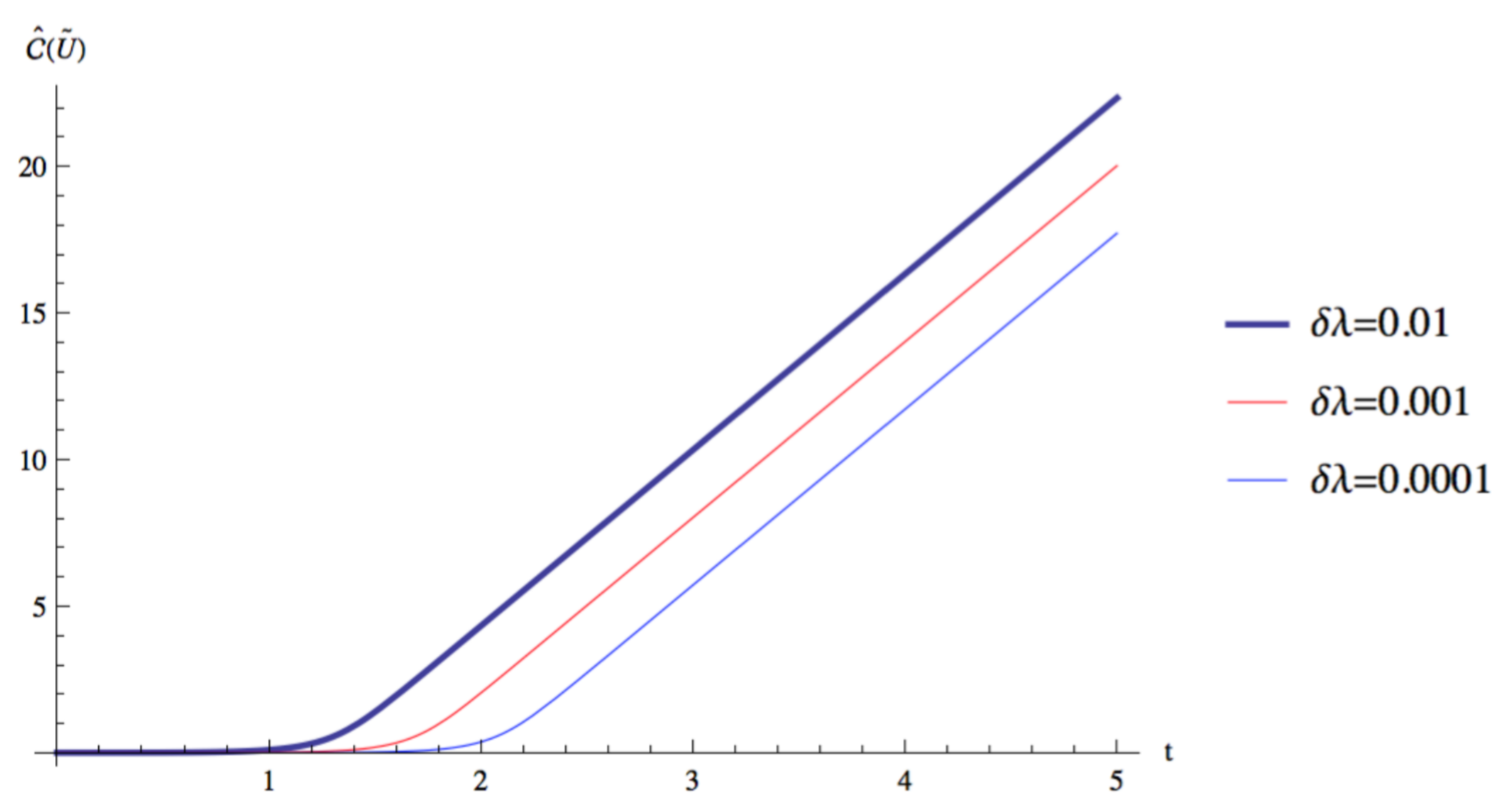} }
\caption{Dependence of $\mathcal{\hat C}(\tilde U)$ on $\delta \lambda$ ($m=1,\lambda=10$)}
\label{newfig}
\end{figure}

We further check the sensitivity of our results to the magnitude of $\delta \lambda.$ We plot the $\mathcal{\hat C}(\tilde U)$ for the inverted oscillator for a fixed value of $\lambda$ but for different $\delta \lambda.$ We find that while the slope of the linear region remains same, the pick up time is sensitive to $\delta \lambda$ as exhibited in Fig.~(\ref{newfig}).\\

We will conclude this section by highlighting the fact that we get the same result regarding diagnosing chaos when we explore the circuit complexity (by correlation matrix method), where both target and reference states are evolved by slightly different Hamiltonians from some other state. The equality between these two complexities was concretely shown in \cite{me1}.
\section{OTOC, Lyapunov Exponent and Scrambling Time}
%%%
The exponential behaviour of the 4-point OTOC has recently emerged as a popular early-time diagnostic for quantum chaos\footnote{An alternative to the OTOC, $F(t) = \langle A^{\dagger}(t)B^{\dagger}(0)A(t)B(0)\rangle$, is the thermally averaged commutator-squared $C(t) = \langle[A(t),B(0)]^{2}\rangle$ with the two being related through $C(t) = 2 - 2\mathrm{Re}(F(t))$. Unless there is an explicit ambiguity, we will refer to them both as the OTOC.}. In \cite{Hashimoto:2017oit} the authors demonstrate explicit calculations of OTOCs for harmonic oscillator. For our model, the OTOC for $x$ and $p$ operators (after reinstating  the factor of $\hslash$) gives \cite{Hashimoto:2017oit}
\begin{figure}[ht] 
\centering
\scalebox{0.30}{\includegraphics{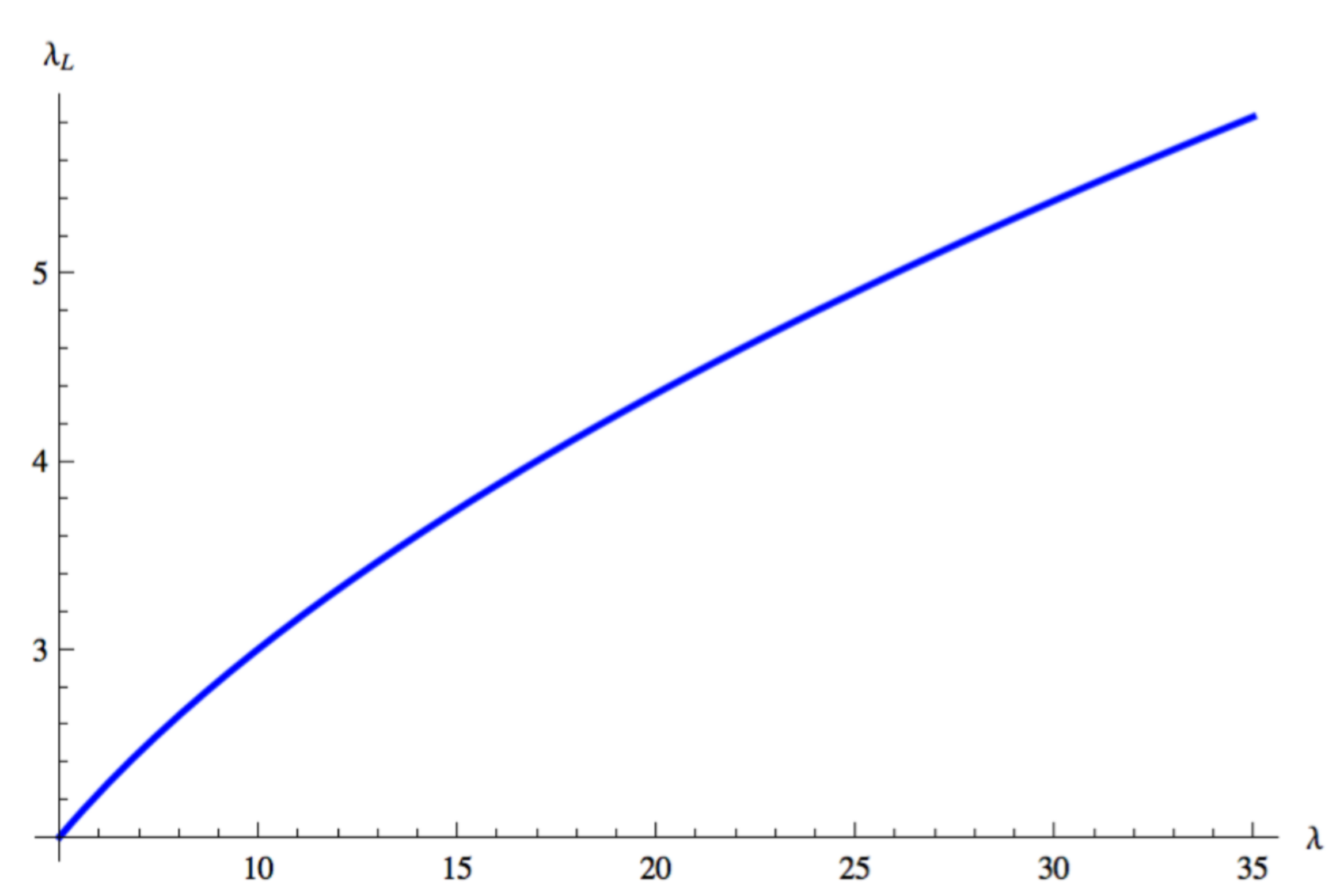} }
\scalebox{0.30}{\includegraphics{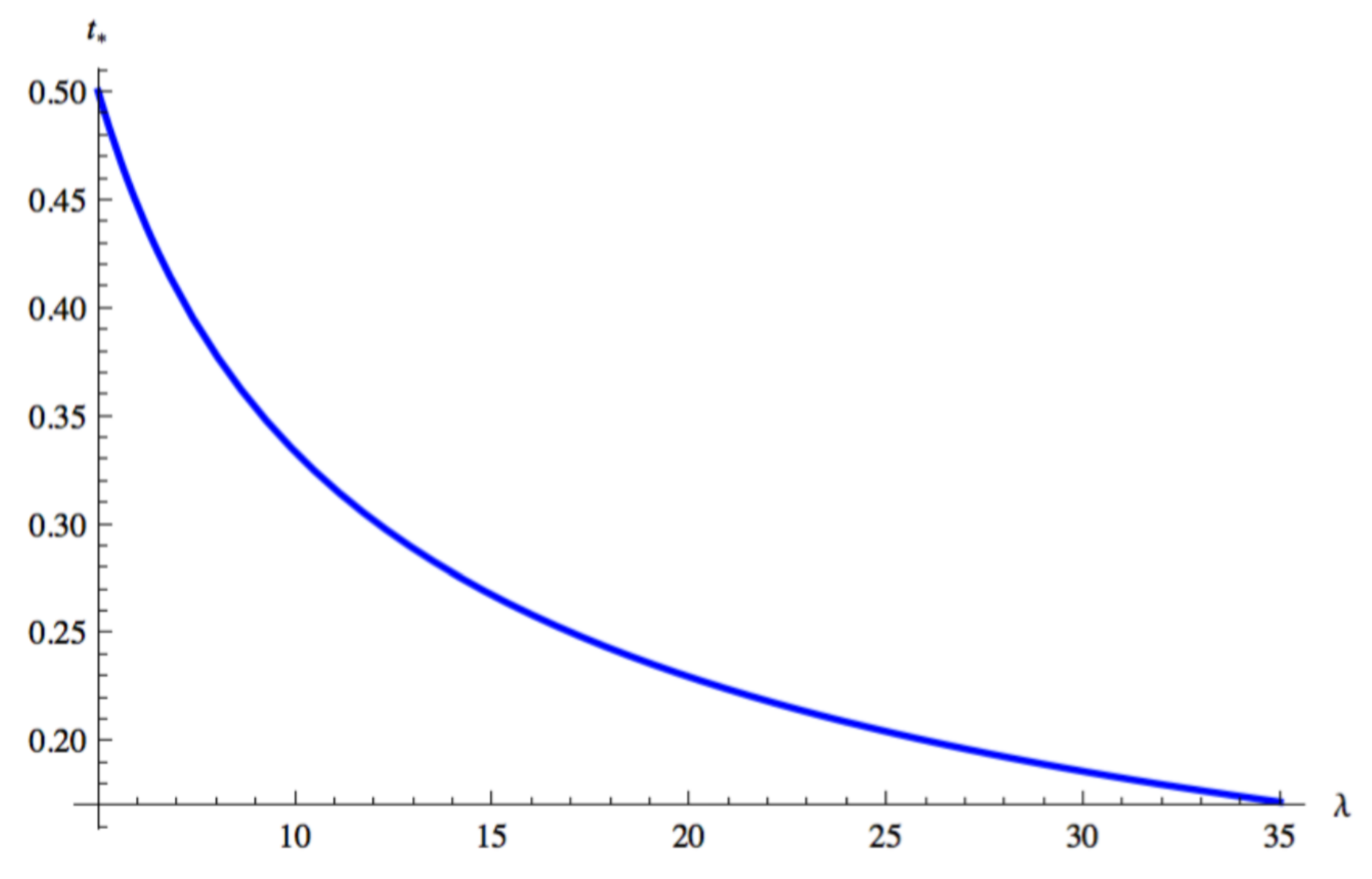} }
\caption{(a)  (a) $\lambda_L$ vs $\lambda$ (m=1). (b) $t_*$ vs $\lambda$ (m=1). }
\label{scram}
\end{figure}
\be  \label{New6}
\langle [x(t), p]^2 \rangle =\hslash^2 \cos^2 \Omega\, t,
\ee
%{\bf AB: This is not OTOC, write it carefully}\\
where $\Omega$ is defined (\ref{Ham}). When $\Omega$ is imaginary, we can write the above expression as an exponential function 
\be \label{New4}
\langle [x(t), p]^2 \rangle  \approx \hslash^2 e^{2\,| \Omega |\, t}+\cdots.
\ee
Rewriting the above expression as $ e^{2\lambda_L (t -t_*)}$, with Lyapunov exponent $\lambda_L$ allows us to immediately read off that for our system, $\lambda_L=|\Omega |$ while the scrambling (or Ehrenfest) time is given by \be \label{New1} t_*= \frac{1}{ \lambda_L}\log \frac{1}{\hslash}.\ee

The $\lambda$ dependence of this time scale (in the units of $\log \frac{1}{\hslash}$) is shown in Fig.~(\ref{scram}b). The nature of the graph is in agreement with Fig.~(\ref{lyap3}b). In fact  from  Fig.~(\ref{scram}b) after doing  a data-fitting we get for the pick-up time,
\be \label{New2}
t_{s}=\frac{4\log(2)}{|\Omega |}.
\ee
In the scale of $\log \frac{1}{\hslash}$ this is  related to scrambling time $t_*$ as, \be  \label{New3} t_{s}=4\log(2) t_{*}.\ee
 Also, the $\lambda$ dependence of the Lyapunov exponent is shown in Fig.~(\ref{scram}a). Again the nature of the graph is in agreement with Fig.~(\ref{lyap3}a). After fitting the data we get for the slope $\phi$ of the linear region of the graph in Fig.~(\ref{lyap3}a),
 \be
 \phi = 2|\Omega|=2\,\lambda_{L}.
 \ee

We have also checked the $m$-dependence of the slope $\phi$ and the pick up time $t_s$  and they are in agreement with the $m$-dependence of $\lambda_L$ and $t_*$ respectively. \\

Before we end this section we would like to highlight the following interesting point. Mathematically, it is evident from equations (\ref{New6}) and (\ref{New4}) that the exponential decay of OTOC is a consequence of a simple analytic continuation. This is due to the change of  $\cos \Omega\, t $ to $\cosh \Omega\, t ,$  when $\Omega$ becomes imaginary. This implies that a simple analytic continuation is essentially capturing the scrambling and chaotic behaviour in this quantum system.
%%%%%
%%%%%
\section{Towards a Field Theory Analysis}
By using the single oscillator model we have illustrated how complexity can capture chaotic behaviour. In this section we will explore a possible field theory model in which the inverted oscillator appears naturally. Consider two free scalar field theories ((1+1)-dimensional $c=1$ conformal field theories) deformed by a marginal coupling. The Hamiltonian is given by
\begin{align}
\begin{split}
H= H_0+H_I= \frac{1}{2}\int dx \Big[\Pi_{1}^2+(\partial_{x}\phi_1)^2+\Pi_{2}^2+(\partial_{x}\phi_2)^2+m^2 (\phi_1^2+\phi_2^2)\Big]+ \lambda \int dx (\partial_{x}\phi_1)(\partial_{x}\phi_2).
\end{split}
\end{align}
We can discretize this theory by putting it on a lattice. Then using the following definitions 
\begin{align}
\begin{split} 
x(\vec n)=\delta \phi(\vec n), \ p(\vec n)=\Pi(\vec n)/\delta, \ \omega=m, \ \Omega=\frac{1}{\delta^2},\,  \hat \lambda=\lambda\, \delta^{-4} \ \text{and}\ \hat m=\frac{m}{\delta},
\end{split}
\end{align}
we get
\begin{align}
\begin{split}
H=\frac{\delta }{2}\sum_{n}&\Big[p_{1,n}^2+p_{2,n}^2+\Big(\Omega^2\,(x_{1,n+1}-x_{1,n})^2+\Omega^2\,(x_{2,n+1}-x_{2,n})^2+\\& \big(\hat m^2( x_{1,n}^2+ x_{2,n}^2)+\hat \lambda\, (x_{1,n+1}-x_{1,n})(x_{2,n+1}-x_{2,n})\Big)\Big].
\end{split}
\end{align}
Next we perform a  series of transformations,
%\begin{align}
%\begin{split}
\bea
x_{1,a} &= &\frac{1}{\sqrt{N}}\sum_{k=0}^{N-1} \exp\Big(\frac{2\pi\,i\,k}{N}\, a\Big)\tilde x_{1,k}, \cr
p_{1,a} &= &\frac{1}{\sqrt{N}}\sum_{k=0}^{N-1} \exp\Big(-\frac{2\,\pi\,i\,k}{N}a\Big)\tilde p_{1, k},\cr
x_{2,a} &= &\frac{1}{\sqrt{N}}\sum_{k=0}^{N-1} \exp\Big(\frac{2\pi\,i\,k}{N}\, a\Big)\tilde x_{2,k},\cr
p_{2,a} &= &\frac{1}{\sqrt{N}}\sum_{k=0}^{N-1} \exp\Big(-\frac{2\,\pi\,i\,k}{N}a\Big)\tilde p_{2, k},\cr
\tilde p_{1,k} &=& \frac{p_{s,k}+p_{a,k}}{\sqrt{2}},\ \tilde p_{2,k}=\frac{p_{s,k}-p_{a,k}}{\sqrt{2}},\cr
\tilde x_{1,k} &=& \frac{x_{s,k}+x_{a,k}}{\sqrt{2}},\ \tilde x_{2,k}=\frac{x_{s,k}-p_{a,k}}{\sqrt{2}},
\eea
that lead to the Hamiltonian
\begin{align}
%\begin{split} 
\label{Ham1}
H=\frac{\delta}{2}\sum_{k=0}^{N-1}&\Big[p_{s,k}^2+\bar \Omega_k^2  x_{s,k}^2+p_{a,k}^2+\Omega_k^2 x_{a,k}^2\Big], 
%\end{split}
\end{align}
where 
\be
\bar \Omega_k^2= \left(\hat m^2+4\,(\Omega^2+\hat \lambda)\,\sin^2\Big(\frac{\pi\,k}{N}\Big)\right), \ \Omega_k^2= \left(\hat m^2+4\,(\Omega^2-\hat \lambda)\,\sin^2\Big(\frac{\pi\,k}{N}\Big)\right). 
\ee
%%%
%\begin{figure}[ht] 
%\centering
%\scalebox{0.4}{\includegraphics{Noscillator}}
%\caption{$\mathcal{C}(\tilde U)$ vs $\lambda$ for $\delta=0.1, m=1,t=20, N=1000.$}
%\label{Noscillator}
%\end{figure}
%%%
It is immediately clear that by tuning the value of $\hat \lambda$, the frequencies $\Omega_k$ can be made arbitrarily negative resulting in coupled inverted oscillators. Note that $\bar \Omega_k$ will be always positive. Therefore, one can view (\ref{Ham1}) as a sum of regular and inverted oscillator for each value of $k$. Now to study the unstable behaviour, the regular oscillator part is not very interesting. Hence, we will simply investigate the inverted oscillator part with the Hamiltonian
\begin{align}
\begin{split} \label{Ham2}
\tilde H (m,\Omega,\hat \lambda)=\frac{\delta}{2}\sum_{k=0}^{N-1}&\left[p_{k}^2+\left(\hat m^2+4\,(\Omega^2-\hat \lambda)\,\sin^2\left(\frac{\pi\,k}{N}\right) \right)x_{k}^2\right].
\end{split}
\end{align}
Note that by tuning $\hat\lambda$ for this Hamiltonian one can get both regular and inverted oscillators. %Now we want to time evolve with the above Hamiltonian. 
At $t=0$ we start with the ground state of $\tilde H (m, \Omega, \hat \lambda=0).$  Then we compute $\mathcal{\hat C}(\tilde U)$ as before by considering two Hamiltonians $\tilde H$ and $\tilde H'$ with two slightly different couplings, $\hat \lambda$ and $\hat \lambda'=\hat \lambda+\delta \hat \lambda,$ where $\delta \hat\lambda$ is small. We get
%%%
 \begin{figure}[ht] 
\centering
\scalebox{0.35}{\includegraphics{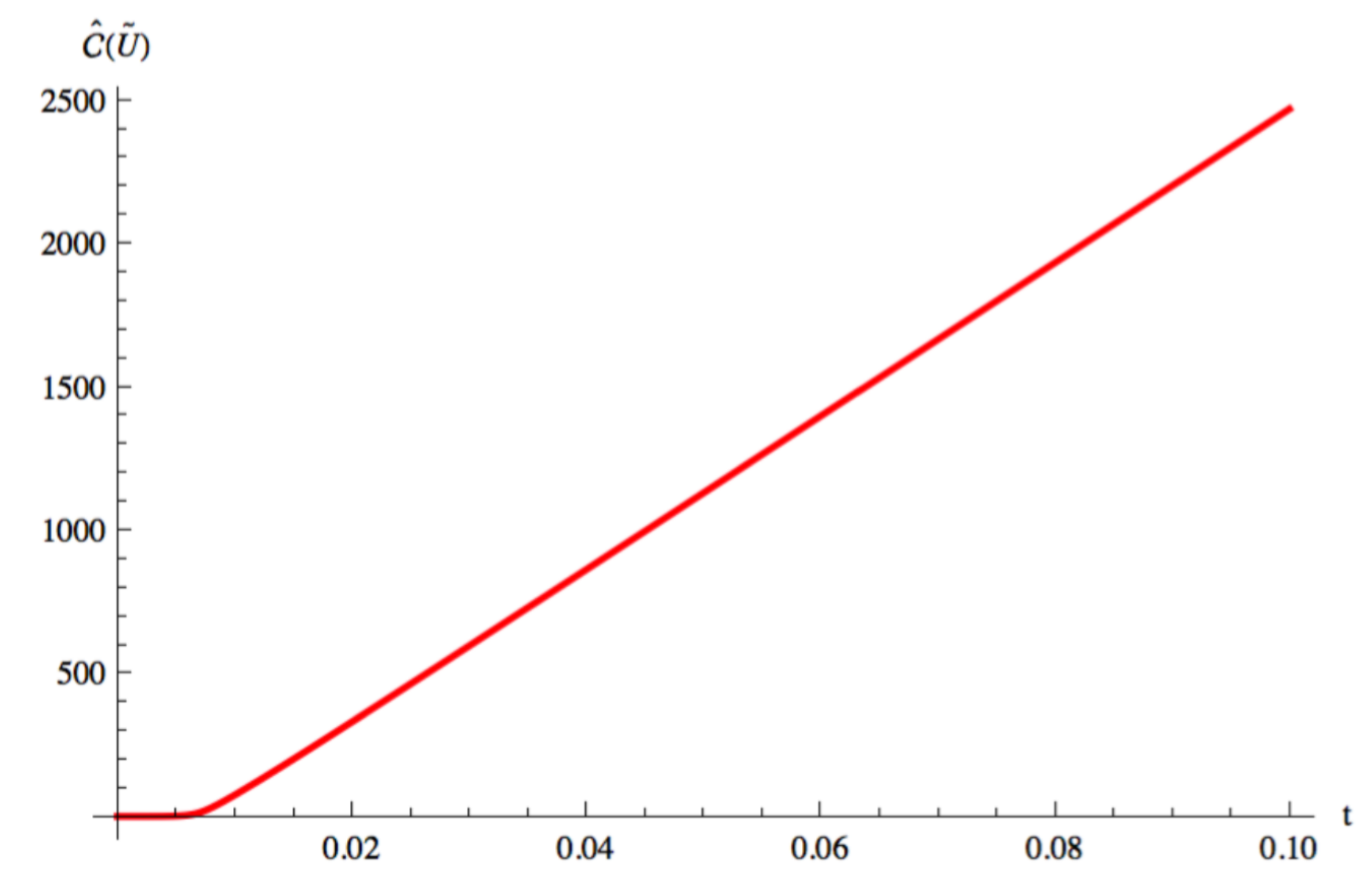}}
\caption{$\mathcal{\hat C}(\tilde U)$ vs time for the Inverted Oscillators ($\delta=0.1, m=1, N=1000, \hat\lambda\delta^2=10, \delta \lambda=0.01$)}
\label{regular}
\end{figure}
%%%
\be 
\mathcal{\hat C}(\tilde U) =\frac{1}{2}\sqrt{\sum_{k=0}^{N-1}\left(\cosh^{-1}\left[\frac{\omega_{r,k}^2+|\hat \omega_k(t)|^2}{2\,\omega_{r,k}\,\text {Re}  (\hat \omega_k(t))}\right]\right)^2},
\ee
where
 \begin{align}
 \begin{split} \hat \omega_k(t)=  i \ \Omega_k' \cot (\Omega_k' t)  + \frac{\Omega_k'^2}{  \sin^2( \Omega_k' t )\left(\omega_k(t) + i\, \Omega_k' \cot (\Omega_k' t)\right)},\ \Omega_k'^2= \hat m^2+4\,(\Omega^2-\hat \lambda-\delta\hat \lambda)\sin^2\Big(\frac{\pi\,k}{N}\Big)
 \end{split}
 \end{align}
 and $\omega_k(t),\ \omega_{r,k}^2$ are defined as,

%and then time evolve it with $\tilde H (m, \Omega, \hat \lambda \neq 0)$. The complexity for this time evolved target state with respect to the ground state of $\tilde H (m, \Omega, \hat \lambda=0)$ is given by a suitable generalization of (\ref{answ})
%\be
%\mathcal{C}(\tilde U)=\frac{1}{2}\sqrt{\sum_{k=0}^{N-1}\left(\cosh^{-1}\left[\frac{\omega_{r,k}^2+|\omega_{k}(t)|^2}{2\,\omega_{r,k}\,\text {Re}  (\omega_k(t))}\right]\right)^2} ,
%\ee
%where
\be
%\begin{split} 
\label{def1}
\omega_k(t)=\Omega_k \left (\frac{\Omega_k-i\,\omega_{r,k} \cot (\Omega_k\, t)}{\omega_{r,k}-i\,\Omega_k \cot (\Omega_k\, t)}\right),
\ee
and
\be
 \omega_{r,k}^2=\hat m^2+4\,\Omega^2\,\sin^2\Big(\frac{\pi\,k}{N}\Big). %\ \Omega_k^2=\hat m^2+4\,(\Omega^2-\hat \lambda)\sin^2\Big(\frac{\pi\,k}{N}\Big).
\ee
  
%\subsection{Regular {\it vs} Inverted Oscillator in a Lattice:}

%Now we compute the complexity. Just as in the single oscillator case, we observe the formation of a cusp as the system switches from regular to inverted oscillator. Fig.~\ref{Noscillator} shows the appearance of cusp. Moreover, we confirmed that the cusp starts to appear at a certain time scale. 
Using our testing method (outlined in Section 4), once again for the inverted oscillator we can immediately read off the scrambling time and Lyapunov exponent from the time evolution of $\mathcal{\hat C}(\tilde U)$ as shown in from Fig.~(\ref{regular}).\par
%Like before, we have also plotted $\mathcal{\hat C}(\tilde U)$ against $\hat \lambda$ for various values of time $t$ \footnote{ In this case the critical value of $\hat \lambda$ denote that particular value of $\hat \lambda$ at which at least one of the $\omega_{k}$ vanishes.}. Although, we have observed  that at critical value of $\hat \lambda$  there is a sharp increase in magnitude of $\mathcal{\hat C}(\tilde U),$ but  we  do not find any onset time  for this case. It seems that the jump in the magnitude of $\mathcal{\hat C}(\tilde U)$ happens for all values of $t.$

%%%%%%%%%%%%%%%%%%%%%%%%%%%%%%%%%%%%%%%%%%%%%%%%%%%%%%%%%%%
\section{Discussion}
%%%%%%%%%%%%%%%%%%%%%%%%%%%%%%%%%%%%%%%%%%%%%%%%%%%%%%%%%%%

In this paper we used a harmonic oscillator model that converts to an inverted oscillator for large coupling of the interaction Hamiltonian. The coupling behaves as a regulator and by tuning it we can switch between regular and inverted regimes. Our motivation was to use this inverted oscillator as a toy model to study quantum chaos. In this context, the regular oscillator serves as a reference system. We developed a new diagnostic for quantum chaos by constructing a particular quantum circuit and computing the corresponding complexity. Our diagnostic can extract equivalent information as the out-of-time-order correlator with the additional feature that complexity can detect when the system switches from regular to the chaotic regime.\\
 
We considered a target state which is first forward evolved and then backward evolved with slightly different Hamiltonians and found that the behaviour for the regular and inverted oscillator are completely different in this case. For the regular oscillator we get some oscillatory behaviour as in \cite{me,me1,meen}. However, for the inverted oscillator we get an exponential type function with two distinct features: for an initial period the complexity is nearly zero, after which it exhibits a steep linear growth. By comparing with the operator product expansion, we discovered the small time scale and slope of the linear portion to be equivalent to the scrambling time and the Lyapunov exponent respectively. To elaborate further, we note that the only difference between the
scrambling time (as in (\ref{New1})) and pick-up time (as in (\ref{New2})) is just  $\mathcal{O}(1)$ (as evident from (\ref{New3})) constant in natural units. Hence, these definitions are capturing the same scrambling physics.\\
 
 To give a proof-of-principle argument for complexity as a chaos diagnostic, we have used the inverted oscillator as a toy model. This is, however, a rather special example and, by no means, a realistic chaotic system. To put complexity on the same footing as, say the OTOC as a probe of quantum chaos will take much more work, with more `realistic' systems like the maximally-chaotic SYK model\footnote{It is worth pointing out that as this manuscript was being prepared, we were made aware of the recent article \cite{Balasubramanian:2019wgd} also advocing for the idea of complexity as a tool for the study of quantum chaotic systems.} and its many variants (see, for example, \cite{Maldacena:2016hyu, Fu:2016vas, Kitaev:2017awl} and references therein) in the (0+1)-dimensional quantum mechanical context, or the MSW class of (1+1)-dimensional (non-maximally) chaotic conformal field theories \cite{Murugan:2017eto}.\\

As a final point of motivation,  we note that by virtue of the recent `complexity=action' \cite{Brown:2015bva} and `complexity=volume' \cite{Roberts:2014isa} conjectures, the computational complexity of  holographic quantum system has a well-defined (if not entirely unambiguous) dual. This opens up tantalising new possibilities in the study of quantum chaos in strongly coupled quantum systems. We leave these issues for future work.

%%%%%%%%%%%%%%%%%%%%%%%%%%%%%%%%%%%%%%%%%%%%%%%%%%%%%%%%%%%%%%%%%%%%%%%%%%%
%%%%%%%%%%%%%%%%%%%%%%%%%%%%%%%%%%%%%%%%%%%%%%%%%%%%%%%%%%%%%%%%%%%%%%%%%%%
%%%%%%%%%%%%%%%%%%%%%%%%%%%%%%%%%%%%%%%%%%%%%%%%%%%%%%%%%%%%%%%%%%%%%%%%%%%
%%%%%%%%%%%%%%%%%%%%%%%%%%%%%%%%%%%%%%%%%%%%%%%%%%%%%%%%%%%%%%%%%%%%%%%%%%%

\section*{Acknowledgements} 
We would like to thank Dario Rosa and Jon Shock for useful comments. AB thanks Aninda Sinha, Pratik Nandy, Jose Juan Fernandez-Melgarejo and Javier Molina Vilaplana for many discussions and ongoing collaborations on complexity. AB also thanks  Department of Physics, University of Windsor, Perimeter Institute and FISPAC group (specially Jose Juan Fernandez-Melgarejo and E. Torrente-Lujan)of Department of Physics, University of Murcia for their warm hospitality during this work.  AB is supported by JSPS Grant-in-Aid for JSPS fellows (17F17023). JM is supported by the NRF of South Africa under grant CSUR 114599. NM is supported by the South African Research Chairs Initiative of the Department of Science and Technology and the National Research Foundation of South Africa. Any opinion, finding and conclusion or recommendation expressed in this material is that of the authors and the NRF does not accept any liability in this regard. Research at Perimeter Institute is supported by the Government of Canada through the Department of Innovation, Science, and Economic Development, and by the Province of Ontario through the Ministry of Research and Innovation.
 
%%%%%%%%%%%%%%%%%%%%%%%%%%%%%%%%%%%%%%%%%%%%%%%%%%%%%%%%%%%%%%%%%%%%%%%%%%%
%%%%%%%%%%%%%%%%%%%%%%%%%%%%%%%%%%%%%%%%%%%%%%%%%%%%%%%%%%%%%%%%%%%%%%%%%%%
%%%%%%%%%%%%%%%%%%%%%%%%%%%%%%%%%%%%%%%%%%%%%%%%%%%%%%%%%%%%%%%%%%%%%%%%%%%
%%%%%%%%%%%%%%%%%%%%%%%%%%%%%%%%%%%%%%%%%%%%%%%%%%%%%%%%%%%%%%%%%%%%%%%%%%%

%%%%%%%%%%%%%%%%%%%%%%%%%%%%%%%%%%%%%%%%%%%%%%%%%%%%%%%%%%%%%%%%%%%%%%%%%%%
%%%%%%%%%%%%%%%%%%%%%%%%%%%%%%%%%%%%%%%%%%%%%%%%%%%%%%%%%%%%%%%%%%%%%%%%%%%
%%%%%%%%%%%%%%%%%%%%%%%%%%%%%%%%%%%%%%%%%%%%%%%%%%%%%%%%%%%%%%%%%%%%%%%%%%%
%%%%%%%%%%%%%%%%%%%%%%%%%%%%%%%%%%%%%%%%%%%%%%%%%%%%%%%%%%%%%%%%%%%%%%%%%%%

\end{document}